\documentclass[12pt,preprint]{aastex}

\usepackage{graphicx}
\usepackage{url}

\newcommand{\rmsub}[1]{\ensuremath{_{\mathrm{#1}}}}
\newcommand{\tento}[1]{\ensuremath{\times 10^{#1}}}
\newcommand{\us}{\ensuremath\mathrm{\mu s}} %microsecond
\newcommand{\ms}{\textrm{ms}} %millisecond
\newcommand{\Gyr}{\textrm{Gyr}} %gigayear
\newcommand{\pssq}{\ensuremath{\mathrm{s^{-2}}}} %per cubic parsec
\newcommand{\m}{\textrm{m}} %meter
\newcommand{\kpc}{\textrm{kpc}} %kiloparsec
\newcommand{\pcubpc}{\ensuremath{\mathrm{pc^{-3}}}} %per cubic parsec
\newcommand{\K}{\textrm{K}} %Kelvin
\newcommand{\GHz}{\textrm{GHz}} %gigahertz
\newcommand{\MHz}{\textrm{MHz}} %megahertz
\newcommand{\mJy}{\textrm{mJy}} %millijanksy
\newcommand{\uJy}{\ensuremath{\mathrm{\mu Jy}}} %microjanksy
\newcommand{\dmu}{\ensuremath{\mathrm{pc\, cm^{-3}}}}
\newcommand{\Msun}{\ensuremath{\mathrm{M_{\Sun}}}} % solar mass
\newcommand{\Lsun}{\ensuremath{\mathrm{L_{\Sun}}}} %solar luminosity
\newcommand{\Rsun}{\ensuremath{\mathrm{R_{\Sun}}}} %solar radius
\newcommand{\clt}{{\it c}} %speed of light

\defcitealias{hrs+07}{HRS07} 	

\title{A New Pulsar in Green Bank Telescope Searches of Ten Globular
  Clusters}

\shorttitle{A New Globular Cluster Pulsar}

\author{Ryan S. Lynch}
\affil{Department of Astronomy, University of Virginia}
\affil{P.O.\ Box 400325, Charlottesville, VA 22904-4325}
\email{rsl4v@virginia.edu}
\and
\author{Scott M. Ransom}
\affil{National Radio Astronomy Observatory}
\affil{520 Edgemont Road, Charlottesville, VA 22903-2454}
\email{sransom@nrao.edu}

\shortauthors{Lynch \& Ransom}


\begin{document}

\begin{abstract}

  We report the results of pulsar searches in ten globular clusters
  using the Robert C.\ Byrd Green Bank Telescope.  One new binary
  millisecond pulsar has been discovered in NGC~5986 with
  $P\rmsub{spin} = 2.6\; \ms$, $P\rmsub{orb} = 1.3\ \mathrm{d}$, and a
  minimum companion mass of $0.16\; \Msun$.  The companion is most
  likely a Helium white dwarf.  Eight of the globular clusters we
  searched have central densities $<10^4\; \Lsun\, \pcubpc$, making
  this a good sample for studying the pulsar content of low density
  clusters.  We find no evidence for pulsars in clusters with very low
  densities ($<10^3\; \Lsun\, \pcubpc$), consistent with theoretical
  predictions.  Null results in many of clusters we searched with
  moderate densities indicate that these systems do not contain a
  bright MSP.  Two clusters in particular, one with very low
  metallicity, stand in contrast to theoretical calculations by
  \citet{ihr+08}.  We also find that three body exchange interaction
  rates calculated by \citep{p96} seem to over predict the pulsar
  content in the clusters we studied.

\end{abstract}

\maketitle

\section{Introduction \label{sec:intro}}

Over the past 23 years, 143 pulsars have been discovered in 27
globular clusters (GCs), the vast majority of which are true
millisecond pulsars (MSPs) with $P\rmsub{s} \lesssim 10\;
\ms$\footnote{See \url{http://www.naic.edu/~pfreire/GCpsr.html} for an
  up-to-date list}.  Many of these MSPs are in exotic binary systems
\citep[e.g.][]{dpm+01,rhs+05,hrs+06,frg07,frb+08}.  This stands in
stark contrast to the population of pulsars found in the Galactic
disk, which are dominated by isolated pulsars with longer spin periods
and a small number of binary MSPs.  These differences have been
attributed to the high stellar densities in GC cores, where exchange
interactions form mass-transferring binaries leading to the recycling
of a ``dead'' pulsar.  These exchange interactions are also
responsible for the exotic binaries described above \citep[see][for
reviews]{c+r05,s08}.  The large number of exchange interactions in GCs
can in turn be traced to the high stellar densities in cluster cores,
and indeed, correlations between density and the number of neutron
stars in GCs have been known for some time.  \citet{p+03} demonstrated
that the core interaction rate, $\Gamma\rmsub{c} \propto
\rho\rmsub{c}^{1.5} r\rmsub{c}^2$, is a good predictor of the number
of low-mass X-ray binaries, which are thought to be the progenitor
systems to many MSPs.  The value of $\Gamma\rmsub{c}/D^2$, where $D$
is the distance and accounts for flux losses, has been used to pick
out the most promising clusters for pulsar searches with great
success.  Terzan 5 \citep{rhs+05,hrs+06} and 47 Tucanae
\citep{clf+00,fck+03} are the two richest clusters for MSPs and have
the highest value of $\Gamma\rmsub{c}/D^2$.  Recent simulations also
show a strong dependence on core density for the number of observable
pulsars \citep{ihr+08}.

Nonetheless, there are exceptions to this rule that have been known
for some time.  \citet{kap+91} discovered two pulsars in the low
density GCs M13 and M53.  Subsequent searches uncovered a substantial
population of MSPs in GCs with $\rho\rmsub{c} \sim 10^3$--$10^4\;
\Lsun\, \pcubpc$.  The totals now stand at five MSPs in M5
\citep{wkm+89,hrs+07}, five in M13 \citep{kap+91,a92,hrs+07}, four in
M3 \citep{hrs+07}, two in NGC~6749 \citep{hrs+07}, one in M53
\citep{kap+91}, and one in M71 \citep{hrs+07}.  These MSPs account for
about 13\% of all known GC pulsars, despite the clusters contributing
only 3\% to the total $\Gamma\rmsub{c}$ of all GCs with pulsars.
However, most of these pulsars were discovered by \citet{hrs+07} using
Arecibo, and without the exquisite sensitivity of this telescope, only
a handful of pulsars would have been found.  Shortly after
\citet{kap+91} announced their results, \citet{jkp92} argued that
pulsar formation in low density systems was enhanced relative to high
density systems, finding $N\rmsub{psr} \propto \rho\rmsub{c}^{0.5}$
(though this came prior to many fruitful searches in higher density
clusters).  \citet{s+p95} suggested that three-body encounters with
primordial binaries could give rise to a small population of pulsars
even in low-density GCs.

We have searched ten clusters using the Robert C.\ Byrd Green Bank
Telescope (GBT).  We selected GCs that had not been searched using the
GBT, focusing primarily on clusters that had low central densities and
large Galactic latitudes (to avoid oversubscribed LST ranges).  Five
clusters (NGC~288, NGC~2298, NGC~6981, NGC~7089, and Pal~12) were
searched using the Parkes radio telescope \citep{pdc+05}, but we
ensured that our searches were at least twice as sensitive.  Since
eight of the clusters have $\rho\rmsub{c} < 10^4\; \Lsun\, \pcubpc$,
this a good sample of deeply searched low density systems.  We have
discovered one new binary MSP in NGC~5986, and we discuss the
implications of non-detections in the other clusters.

\section{Survey Details \label{sec:survey}}

All our searches were carried out using the GBT and the Green Bank
Ultimate Pulsar Processor (GUPPI) back-end \citep{drd+08}.
Observations were made at a central frequency of $2\; \GHz$, with
$800\; \MHz$ of bandwidth broken into 2048 frequency channels,
although persistent radio frequency interference (RFI) reduces the
usable bandwidth to $600\; \MHz$.  We used sampling times of
$40.96$--$64\; \us$.  Total system temperatures were typically
$24$--$28\; \K$.  The contribution from the Galactic background was
estimated by scaling the values from \citet{hss+82} to $2\; \GHz$
assuming a spectral index of $-2.6$, though this was usually $\lesssim
1\; \K$ except for clusters at very low Galactic latitudes.
The limiting flux density of each search was taken to be
\begin{eqnarray}
  S\rmsub{\nu,min} = \frac{\beta \xi\, T\rmsub{tot}}
                           {G \sqrt{n\rmsub{pol} \Delta \nu\, t\rmsub{int}}}
                     \sqrt{\frac{W}{W - P}}
\end{eqnarray}
where $\beta = 1.3$ accounts for digitization losses and $\xi = 10$ is
our limiting S/N. $T\rmsub{tot}$ is the total sky and system
temperature, $G = 1.9$ is the telescope gain, $n\rmsub{pol} = 2$ is
the number of summed polarizations, $\Delta \nu$ is the usable
bandwidth, and $t\rmsub{int}$ is the integration time.  $P$ is the
pulse period and $W$ is the observed pulse width, which is a
combination of the intrinsic pulse width, instrumental time
resolution, and broadening of the pulse due to dispersive smearing and
scattering.  This calculation does not take into account sensitivity
losses due to RFI or Doppler smearing of the pulse due to binary
acceleration.  However, we observed very little RFI in our data, and
since we performed acceleration searches (see below), only pulsars in
very tight binaries should have dropped below our detection threshold.
Integration times and approximate limiting flux densities can be found
in Table \ref{table:gcs}, along with some other properties of each
cluster.

To test the accuracy of our limiting flux density estimates, we
calculated the integration time necessary to just detect the lone
pulsar we discovered in NGC~5986, and then blindly searched this
portion of our original data.  The pulsar was indeed detected with
with an appropriate S/N.  We are therefore confident that our reported
limits are accurate.

Data were reduced using the
\texttt{PRESTO}\footnote{\url{http://www.cv.nrao.edu/~sransom/presto/}}
software suite \citep{rem02}.  After removing RFI, de-dispersed time
series were created, starting with a minimum dispersion measure (DM)
of $10\; \dmu$ and increasing in steps of $0.1\; \dmu$ to twice the DM
predicted by the NE2001 model \citep{c+l02}.  Acceleration techniques
were used to search for periodic signals from isolated and binary MSPs
and single pulse searches were used to look for bright transient
emission.  To search for highly accelerated pulsars, we performed
acceleration searches on $\sim 10$--$20$ minute subsets of each
observation.  We searched up to a maximum acceleration of 800 Fourier
bins in all acceleration searches.  Candidate pulsars were visually
inspected and grouped into likely pulsars or random noise and RFI.  We
were able to observe some clusters twice, which allowed us to confirm
or reject candidate pulsars quickly.

\section{The Binary Pulsar in NGC~5986 \label{sec:NGC5986A}}

\begin{figure}
\centering
\includegraphics[width=6in]{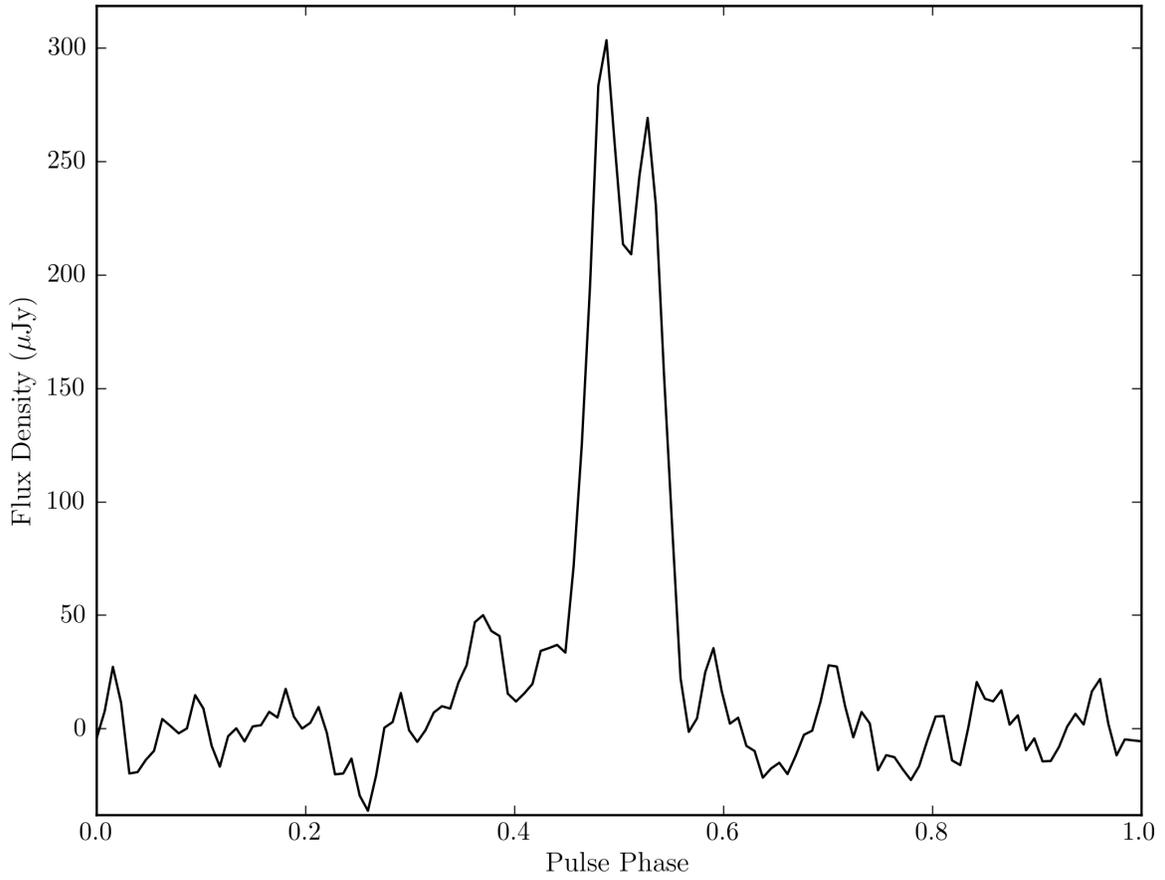}
\caption{The flux-calibrated pulse profile of NGC~5986A at an
  observing frequency of $2\; \GHz$ ($S\rmsub{\nu} = 21\; \uJy$).
  \label{fig:profile}}
\end{figure}

We discovered one new pulsar in NGC~5986, J1546$-$3747A, hereafter
NGC~5986A (Figure \ref{fig:profile}).  The pulsar was discovered in
our acceleration searches with an acceleration of $-26$ Fourier bins,
or $-1.1\; \m\, \pssq$, indicating that it is in a binary system.  We
conducted three follow-up observations to determine the orbit of
NGC~5986A and to search for more pulsars using GUPPI in a coherent
de-dispersion mode at $820\; \MHz$, $1.5\; \GHz$, and $2\; \GHz$.  The
pulsar was detected in all three follow-up observations.  We searched
the $820\; \MHz$ and $2\; \GHz$ observations, but found no new
pulsars.  The $1.5\; \GHz$ observations were heavily contaminated with
RFI, making a blind search impractical.

\begin{figure}
\centering
\includegraphics[width=6in]{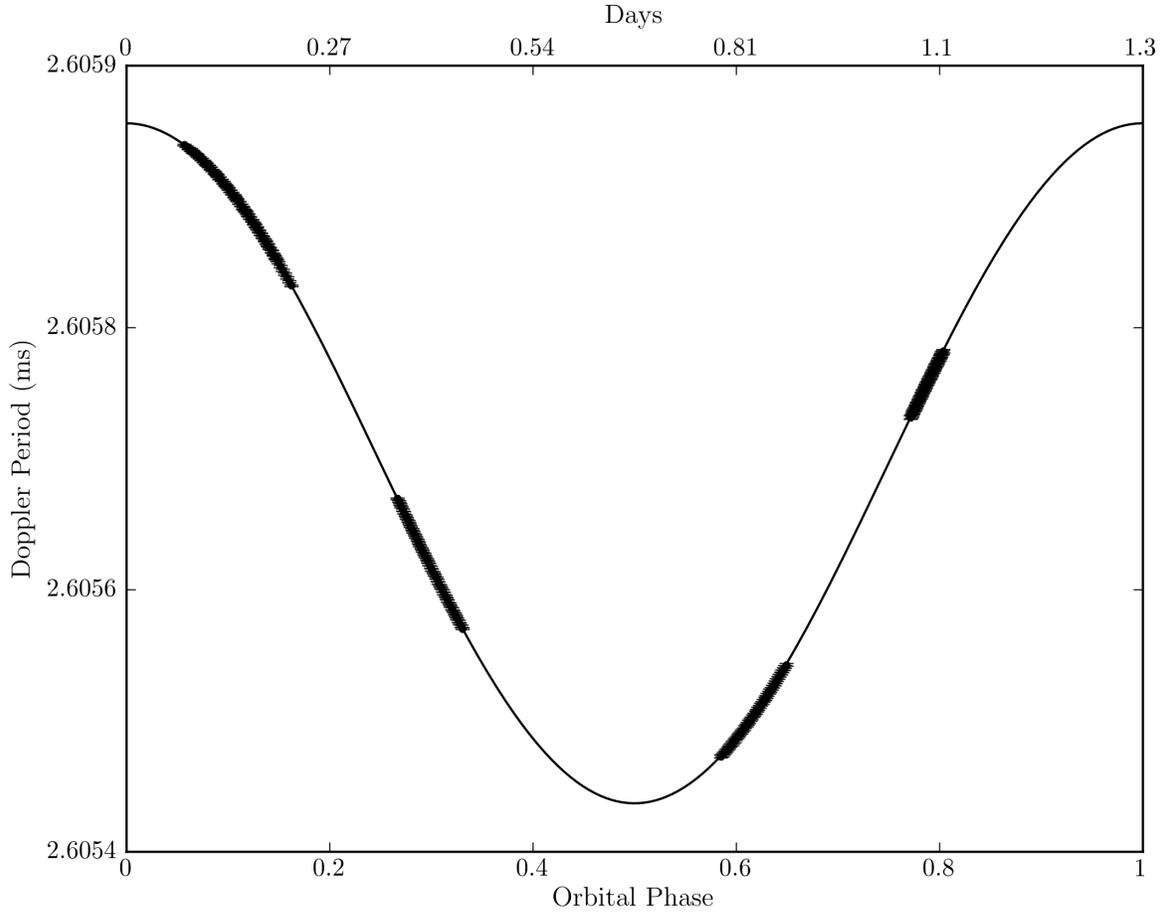}
\caption{The Doppler modulated period of NGC~5986A, clearly showing
  the effects of binary acceleration and a circular orbit.  The thick
  lines show the observed period during our observations, and the thin
  line is the predicted period based on our orbital solution.
  \label{fig:sine}}
\end{figure}

While these observations are not sufficient to obtain a full timing
solution for NGC~5986A, they have allowed us to accurately determine
the orbital parameters (see Table \ref{table:NGC5986A} and Fig.
\ref{fig:sine}).  The orbit is circular, with a period of $\sim
1.3~\mathrm{days}$ and projected semi-major axis of $\sim 0.6\;
\Rsun$.  The companion mass limit is $0.16\; \Msun$ assuming a pulsar
mass of $1.4\; \Msun$.  Given a distribution of orbital inclinations
that is flat in $\cos{i}$, an edge-on orbit is statistically most
likely, so the true companion mass is probably close to this limit.
These binary characteristics are similar to those of MSPs with Helium
white dwarf companions that are found in both GCs and the Galactic
disk.  A main-sequence companion seems unlikely, as such systems have
only been found in dense GCs \citep{egc+02,pdm+03,hrs+06}.
Furthermore, we see no evidence for eclipses, although we lack
coverage near conjunction when eclipses would be most likely.  Systems
similar to NGC~5986A have been found in other GCs of comparable
density\footnote{\url{http://www.naic.edu/~pfreire/GCpsr.html}}.  If a
precise position can be measured through future pulsar timing, a
search for an optical counterpart may be feasible.

Rough flux density estimates were obtained by assuming that the
off-pulse RMS noise levels are described by the radiometer equation,
\begin{eqnarray}
  \sigma = \frac{T\rmsub{tot}}
                {G \sqrt{n\rmsub{pol} \Delta \nu\, t\rmsub{int}}}.
\end{eqnarray}
We find $S_{\nu} = 21$, $27$, and $45\; \uJy$ at $2\; \GHz$, $1.5\;
\GHz$, and $820\; \MHz$, respectively, which yields a spectral index
of $\alpha \sim -0.9$, flatter than the canonical value of $-1.7$.

\section{The Pulsar Content of Low-Density Clusters
  \label{sec:low_dens}}

We searched ten clusters for pulsars, five of which have $10^3 <
\rho\rmsub{c} < 4 \times 10^4\; \Lsun\, \pcubpc$, but found only one
MSP.  This stands in contrast to previous surveys \citep[and
references therein, hereafter \citetalias{hrs+07}]{hrs+07} that found
several pulsars in low density clusters.  We therefore consider what
factors may have led to our null results.

Fig. \ref{fig:sensitivity} shows the approximate limiting luminosity
of our searches, where $L\rmsub{min} = S\rmsub{min} D^2$, along with
the $1.4\; \GHz$ pseudo-luminosities of pulsars with reliable flux
density measurements from \citetalias{hrs+07}, scaled to $2\; \GHz$
using a spectral index of $-1.7$ (NGC~5986A is also shown).  It is
immediately obvious that \citetalias{hrs+07} were more sensitive than
our searches.  We can attribute this to two factors.  The first is
that all the GCs from \citetalias{hrs+07} had $D < 8\; \kpc$, with the
exception of M53 ($D = 17.8\; \kpc$).  By comparison, only two
clusters in our survey have $D < 8\; \kpc$, while most have $D > 10\;
\kpc$.  The second factor is that \citetalias{hrs+07} were able to
reach lower limiting flux densities using Arecibo than we could with
the GBT.

Despite these factors, 3/10 of our searches should have been sensitive
to the brightest $\sim 53\%$ of a population similar to the
\citetalias{hrs+07} pulsars, and 8/10 of our searches should have been
sensitive to the brightest two pulsars.  Furthermore, four out of five
of the clusters with $\rho\rmsub{c} > 10^3\; \Lsun\, \pcubpc$ were
among the five most sensitively searched GCs in our sample.  In other
words, we achieved good sensitivity in the most promising clusters.
As discussed in \S\ref{sec:survey}, we are confident that our limiting
flux densities are accurate unless there are pulsars in extremely
tight binaries.  However, since \citetalias{hrs+07} used the same
search procedure as we have, they suffered from the same bias, so we
do not believe this is a good explanation for the lack of pulsars in
our sample.  While it is possible that some of the \citetalias{hrs+07}
pulsars could have a steeper spectral index than the assumed value of
$-1.7$, we would also expect some to have flatter spectra, making them
easier to detect at $2\; \GHz$.  It therefore seems likely that most
of the clusters we searched lack a bright MSP.

We now turn to the theoretical results of \citet{ihr+08}, who predict
the number of neutron stars in different types of GCs through Monte
Carlo simulations.  They simulated a variety of clusters, including
those with $\log{(\rho\rmsub{c}/\pcubpc)} = 3, 4, 5, 6$.  We began by
empirically fitting a simple exponential to the results of
\citet{ihr+08} of the form
\begin{eqnarray}
n\rmsub{psrs} = a\, e^{\log{\rho\rmsub{c}}} + b
\end{eqnarray}
with $a = 0.041$ and $b = -0.82$, and where $n\rmsub{psrs}$ is the
number of pulsars formed per $2 \times 10^{5}\; \Msun$ (as per the
\citet{ihr+08} simulations).  We explored other functional forms but
chose this one for its goodness of fit and simplicity.  We then
calculated the expected number of pulsars in our clusters based upon
their central densities and masses\footnote{Masses were taken from
  photometric models by O.~Y.\ Gnedin;
  \url{http://www.astro.lsa.umich.edu/~ognedin/gc/vesc.dat}}.  Next,
we calculated how many pulsars would lie above our limiting flux
densities (at $2\; \ms$), given a power-law distribution, $dN(L)
\propto L^{-1}\, dL$.  We assumed a maximum luminosity equal to the
brightest cluster pulsars, (about $250\; \mJy\, \kpc^2$ at $2\; \GHz$)
and a minimum luminosity of $0.16\; \mJy\, \kpc^2$ (obtained by
scaling the typically assumed lower limit of $0.3\; \mJy\, \kpc^2$ at
$1.4\; \GHz$ to $2\; \GHz$).  We calculated upper and lower estimates
in the same way, fitting to the maximum and minimum expected pulsars
as defined by the errors quoted in \citet{ihr+08}.

Based on these calculations, we expect that about nine pulsars should
have been detected in our sample, with upper and lower estimates of 16
and three pulsars, respectively.  Our results are not very sensitive
to our choices of $L\rmsub{max}$ and $L\rmsub{min}$ since our limiting
luminosities are above the $L\rmsub{min}$ cutoff.  Choosing a flatter
slope for the luminosity distribution only increases the number of
potentially observable pulsars.  As a check on our method, we
performed the same analysis using all the clusters searched by
\citetalias{hrs+07}.  M15 (a dense, massive cluster with eight known
pulsars) was an outlier in these calculations, with far more pulsars
predicted than are observed; when it is excluded from the analysis, we
predict $6^{+15}_{-6}$ observable pulsars in the \citetalias{hrs+07}
sample.  Twenty have been discovered (excluding the eight in M15),
which is consistent with this upper limit.

\citet{p96} also gave birth rates for pulsars in GCs for different
formation mechanisms, the most relevant in our density regime being
three-body exchanges.  The birth rates are given per neutron star, so
we once again use the results of \citet{ihr+08} for an estimate of the
total neutron star content of our clusters.  We also assume an MSP
lifetime of $\sim 10\; \Gyr$.  Based on these theoretical models, we
should have been sensitive to $\sim 1$--$2$ pulsars each in NGC2298,
Terzan 9, and NGC6981, about ten in Pal 6, about two dozen in NGC5986,
$\sim 50$ in M92, and $\sim 70$ in NGC7089.

\section{Discussion \label{sec:discussion}}

\begin{figure}
\centering
\includegraphics[width=6in]{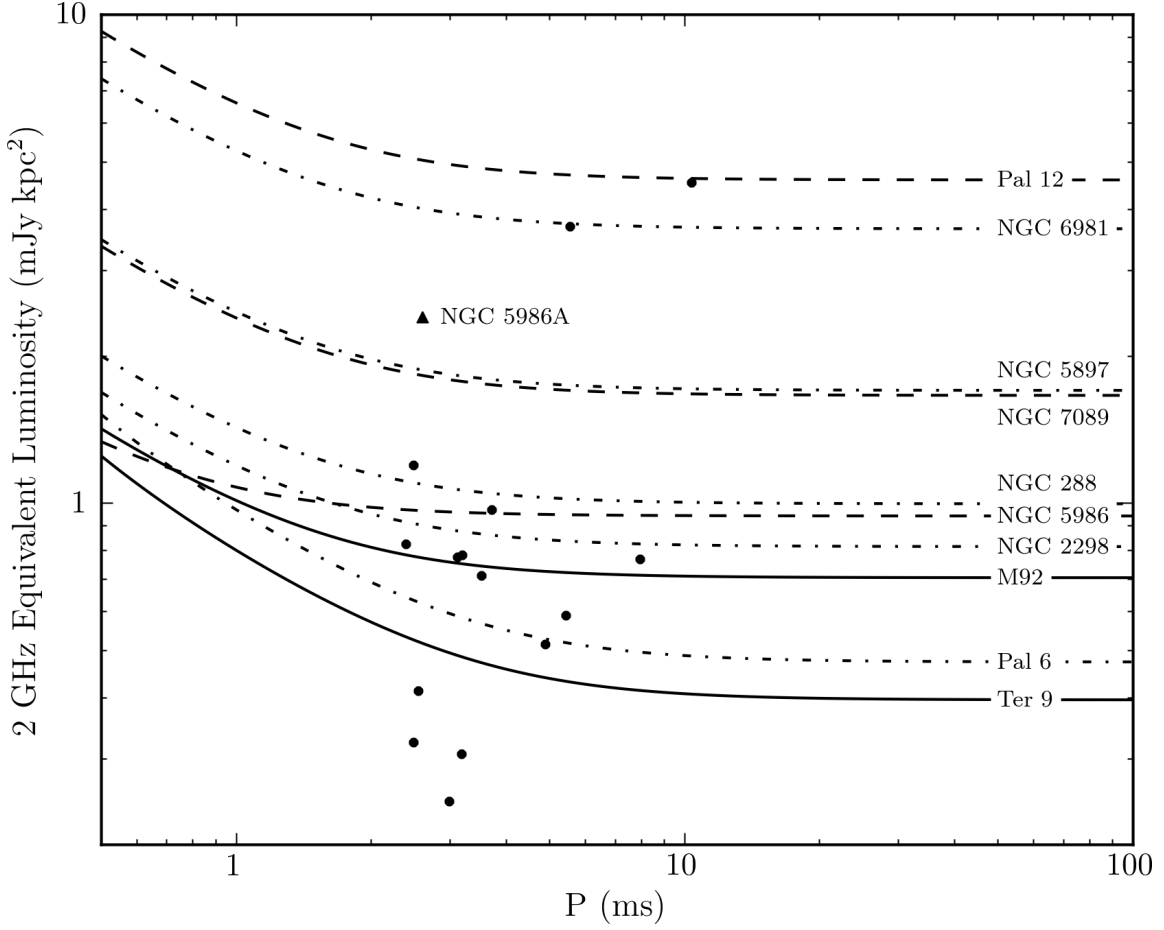}
\caption{Comparison of the limiting luminosity of our searches and the
  luminosities of pulsars with reliable flux densities from
  \citet{hrs+07}, as well as NGC~5986A.  Luminosities of the known
  MSPs were scaled from $1.4\; \GHz$ to $2\; \GHz$ using a spectral
  index of $-1.7$.  Solid lines indicate clusters with $\rho\rmsub{c}
  > 10^4\; \Lsun\, \pcubpc$, dashed lines indicate $10^3$--$10^4\;
  \Lsun\, \pcubpc$, and dot-dashed lines indicate $< 10^3\; \Lsun\,
  \pcubpc$.
  \label{fig:sensitivity}}
\end{figure}

We predict that $9^{+7}_{-3}$ pulsars should have been observable in
our sample based on the results of \citet{ihr+08}.  This prediction
seems high given our single detection in NGC~5986 (we note that this
cluster was predicted to contain one observable pulsar based on the
high estimates of \citet{ihr+08}).  In addition, nearly all of the
potentially observable MSPs come from M92 and NGC~7089.  This
assertion is supported by the fact that these two clusters have the
highest value of $\Gamma\rmsub{c}/D^2$ in our sample.  As can be seen
from Fig.  \ref{fig:sensitivity}, M92 in particular was searched
fairly deeply, so this null result is particularly interesting.  One
parameter that was not taken into account in our simulations was
metallicity.  Surveys of Galactic and extra-galactic GCs have found
$\sim 3$ times more bright LMXBs in metal-rich clusters (defined by
$\mathrm{[Fe/H]} < -1$) \citep{s+07,j+04,mrf+04,kmz02,bpf+95} and it
is reasonable to believe that this effect may also manifest itself in
the MSP population.  We note that eight of the clusters we studied
have $\mathrm{[Fe/H]} < -1$, and that M92 in particular has the second
lowest metallicity of any Milky Way GC---the most metal poor is M15,
which was greatly under-populated in MSPs based on the calculations
discussed in \S\ref{sec:low_dens}.  While a full treatment of
metallicity effects is beyond the scope of this paper, our results may
suggest that MSPs are less likely to be found in low metallicity
clusters.

We find that the results of \citet{p96} drastically over-predict the
number of MSPs that should be observable in our study.  One possible
explanation for this discrepancy is that our assumed MSP lifetime of
$10\; \Gyr$ is too long.  An order of magnitude decrease in the
lifetime would yield numbers that are closer to our observed results.
However, based on observed spin down rates, MSPs should be very long
lived.  The other parameters in our calculations were the number of
primordial neutron stars (which we based off of \citet{ihr+08}) and
the rate of MSP formation through exchange encounters calculated by
\citet{p96}.  It therefore seems that one or both of these inputs is
too large for clusters of low to moderate density.  \citet{p96} over
predicts the number of observable MSPs in higher density clusters as
well, so it seems that these rates may not be applicable when trying
to predict the number of MSPs in a given cluster.

We thus draw the following conclusions.  GC pulsar searches are still
sensitivity limited, and this necessarily makes any statements about
the total population of cluster pulsars a matter of extrapolation.
Nevertheless, our results show no evidence for a population of pulsars
in very low density systems ($<10^3\; \Lsun\, \pcubpc$), which is
consistent with theoretical predictions \citep{ihr+08}.  Even given
our sensitivity limits, though, it appears that clusters with
densities $10^3$--$10^4\; \Lsun\, \pcubpc$ are not as efficient at
forming MSPs as the results of \citetalias{hrs+07} imply.  A full
treatment of the effects of metallicity may shed more light on this
discrepancy.  We find evidence that low (and high) density GCs are
either not as efficient at forming neutron stars or recycling them
into MSPs through exchange interactions as earlier calculations by
\citet{p96} imply.

We would like to thank an anonymous referee whose helpful comments
improved the quality of this manuscript.  R.\ Lynch was supported
through the GBT Student Support program and the National Science
Foundation grant AST-0907967 during the course of this work.  The
National Radio Astronomy Observatory is a facility of the National
Science Foundation operated under cooperative agreement by Associated
Universities, Inc.

\begin{deluxetable}{lcccccccccc}
\rotate
\centering
\tabletypesize{\small}
\tablewidth{0pt}
\tablecolumns{11}
\tablecaption{Globular Clusters Included in Survey \label{table:gcs}} 
\tablehead{\colhead{ID} & \colhead{$\ell$} & \colhead{$b$} &
  \colhead{$D$} & \colhead{Predicted DM} & \colhead{$r\rmsub{c}$} &
  \colhead{$\log{\rho\rmsub{c}}$} & \colhead{$\Gamma\rmsub{c}/D^2$} &
  \colhead{[Fe/H]} & \colhead{$t\rmsub{obs}$} & 
  \colhead{Approximate $S\rmsub{\nu,min}$} \\
  \colhead{} & \colhead{(deg)} & \colhead{(deg)} & \colhead{(\kpc)} &
  \colhead{(\dmu)} & \colhead{(arcmin)} & \colhead{($\Lsun\,
    \pcubpc$)} & \colhead{(\%\tablenotemark{a})} & \colhead{} &
  \colhead{(hr)} & \colhead{($\uJy$)}}
\startdata
NGC~288	 & 152.30 & $-89.38$ & 8.9  & 28   & 1.35 & 1.78 & 0.03  & $-1.32$ & 0.95 & 11 \\
NGC~2298 & 245.63 & $-16.00$ & 10.8 & 85   & 0.31 & 2.90 & 0.08  & $-1.92$ & 3.1  & 6  \\
NGC~5897 & 342.95 & $+30.29$ & 12.5 & 61   & 1.40 & 1.53 & 0.01  & $-1.90$ & 1.3  & 10 \\
NGC~5986 & 337.02 & $+13.27$ & 10.4 & 92.2 & 0.47 & 3.41 & 1.1   & $-1.59$ & 2.2  & 7  \\
M92      & 68.34  & $+34.86$ & 8.3  & 43   & 0.26 & 4.30 & 7.0   & $-2.31$ & 1.4  & 9  \\
Pal~6    & 2.10   & $1.78$   & 5.8  & 397  & 0.66 & 3.46 & 2.5   & $-0.91$ & 1.2  & 15 \\
Terzan~9 & 3.61   & $-1.99$  & 7.1  & 422  & 0.03 & 4.42 & 0.14  & $-1.05$ & 2.5  & 10 \\ 
NGC~6981 & 35.16  & $-32.68$ & 17.0 & 55   & 0.46 & 2.38 & 0.03  & $-1.42$ & 1.0  & 11 \\ 
NGC~7089 & 53.37  & $-35.77$ & 11.5 & 46   & 0.32 & 4.00 & 3.8   & $-1.65$ & 1.0  & 11 \\
Pal~12   & 30.51  & $-47.68$ & 19.0 & 40   & 0.02 & 3.64 & 0.004 & $-0.85$ & 1.0  & 11 \\
\tableline
47 Tucanae\tablenotemark{b} & 305.89 & $-44.89$ & 4.5  & 24.4  & 0.36 & 4.88 & 100 & $-0.72$ & \nodata & \nodata \\
\enddata
\tablenotetext{a}{Normalized to 47 Tucanae.}
\tablenotetext{b}{47 Tucanae has been included so that the properties
  of our sample can be compared to a rich GC.}
\tablecomments{Cluster parameters are taken from the 2010 version of
  the Harris catalog
  (\url{http://physwww.physics.mcmaster.ca/~harris/mwgc.dat}).  DM
  estimates for all clusters except NGC~5986 and 47 Tucanae are from the
  NE2001 model \citep{c+l02}.}
\end{deluxetable}

\begin{deluxetable}{lr}
\centering
\tablewidth{0pt}
\tablecolumns{2}
\tablecaption{Timing Derived Parameters of NGC~5986 \label{table:NGC5986A}}
\tablehead{\colhead{Parameter} & \colhead{Value}}
\startdata
$P$ (ms)                 & 2.6056722466(1)          \\
Reference Epoch (MJD)    & 51900                    \\
DM (\dmu)                & 92.17(4)                 \\
$P\rmsub{b}$ (d)         & 1.3467116(2)             \\
$a \sin{i}/\clt$ (s)     & 1.38325(4)               \\
$T_0$ (MJD)              & 55355.67467(2)           \\
$e$                      & $< 8 \tento{-6}$         \\
$\omega$ (deg)           & \dots                    \\
Mass Function (\Msun)    & 0.0016                   \\
$M\rmsub{c,min}$\tablenotemark{a} (\Msun) & 0.16    \\
$N\rmsub{TOAs}$          & 79                       \\
\enddata
\tablenotetext{a}{Assuming $M\rmsub{psr} = 1.4\; \Msun$}
\tablecomments{All parameters assume that the pulsar is at the cluster
  center, $\alpha = 15$:$46$:$03.44$, $\delta = -37$:$47$:$10.1$
  \citep{s+w86}.  The eccentricity limit was calculated according to
  $e\rmsub{lim} = \delta t\: (a \sin{i}/\clt)^{-1}$ where $\delta t$ is
  our timing precision \citep{p92}.  The orbital solution was obtained
  using the DE405 Solar System ephemeris and the UTC(NIST) time
  standard.}
\end{deluxetable}

\end{document}